\documentclass[aps,prl,twocolumn,groupedaddress,showpacs,showkeys]{revtex4}
\usepackage{graphics}

\begin{document}

\title{Meson model for $f_0(980)$ production in peripheral
  pion-nucleon reactions }

\author{F.P. Sassen}
\author{S. Krewald}
\author{J. Speth}
\affiliation{Institut f\"ur Kernphysik, \\
Forschungszentrum J\"ulich GmbH, 52425 J\"ulich,
Germany}
\author{A.W. Thomas}
\affiliation{Special Research Centre for the Subatomic 
Structure of Matter\\
University of Adelaide, Adelaide 5005, Australia}

\date{\today}

\begin{abstract}
The J\"ulich  model for $\pi\pi$-scattering, based on an effective meson-meson Lagrangian is applied to the 
analysis of the $S$-wave production
amplitudes derived from the BNL E852 experiment $\pi^- p \rightarrow
\pi^0 \pi^0 n$ for a pion momentum of 18.3 GeV and the GAMS
experiments performed at $38\textrm{ GeV}$ and $100\textrm{ GeV}$. The unexpected strong dependence
of the S-wave partial wave amplitude on the momentum transfer between
the proton and neutron in the vicinity of the $f_0(980)$ resonance
is explained in our analysis as interference effect between the
resonance and the the non-resonant background.

\end{abstract}



\maketitle

Meson spectroscopy in the scalar-isoscalar channel has received increasing 
interest motivated by the search for non-$q\bar{q}$ mesons, such as
glueballs\cite{Amsler:1998up}.
The large number of experimentally observed $0^{++}$resonances suggests that some of those
resonances
may  have a more complicated structure than the conventional 
$q\bar{q}$ structure \cite{Amsler:2002ey,Abele:2001pv}. 
The $f_0(980)$ has been a candidate for a non-$q\bar{q}$ meson for
more than two decades\cite{Jaffe:1977ig,Weinstein:1990gu,Lohse:1990ew,Barnes:1985cy,Pichowsky:2001qe,Oller:1998ng,Achasov:1998pu}.

Recently, the scalar-isoscalar $\pi\pi$ partial wave amplitudes have been
deduced from two pion interaction obtained via the
charge-exchange reaction $\pi^- p \rightarrow \pi^0 \pi^0 n$
measured for incident pion momentum of 18.3 GeV by the
E852 collaboration at the Brookhaven National Laboratory\cite{Gunter:2000am}.
 In the
vicinity of the invariant two-pion mass $m_{\pi \pi}=980$ MeV, a
peculiar behavior of the $S$-wave amplitude has been observed.
 Such an effect has also previously been reported
by the GAMS collaboration for a beam momentum of 38 GeV \cite{Alde:1995jj}.
 While for
small momentum transfers between the proton and the neutron $( -t < 0.1\;\textrm{GeV}^2)$
the scalar amplitudes show a dip around 1 GeV, a sharp peak is seen
at the same energy for large momentum transfers $( -t > 0.4\;\textrm{GeV}^2 )$.

This observation has been interpreted as evidence for a hard component
in the $f_0(980)$ which would make the interpretation of this scalar
meson
as a $K\bar{K}$ molecule unconvincing
\cite{Klempt:2000ud,Kondashov:1998uh,Anisovich:1995jy,Anisovich:2002us}.
Here we want to show that the strong dependence of the
$f_0(980)$-production on the momentum transfer between the proton and
the neutron is not in contradiction with a strong $K\bar{K}$
contribution to the $f_0(980)$.
 Actually we will show in the following that this $t$-dependence is
 due to the interference between the resonance structure and the non-resonant background
 and does not depend on the detailed structure of the $f_0(980)$.

\begin{figure}
\resizebox{0.48\textwidth}{!}{%
\rotatebox{90}{
 \includegraphics{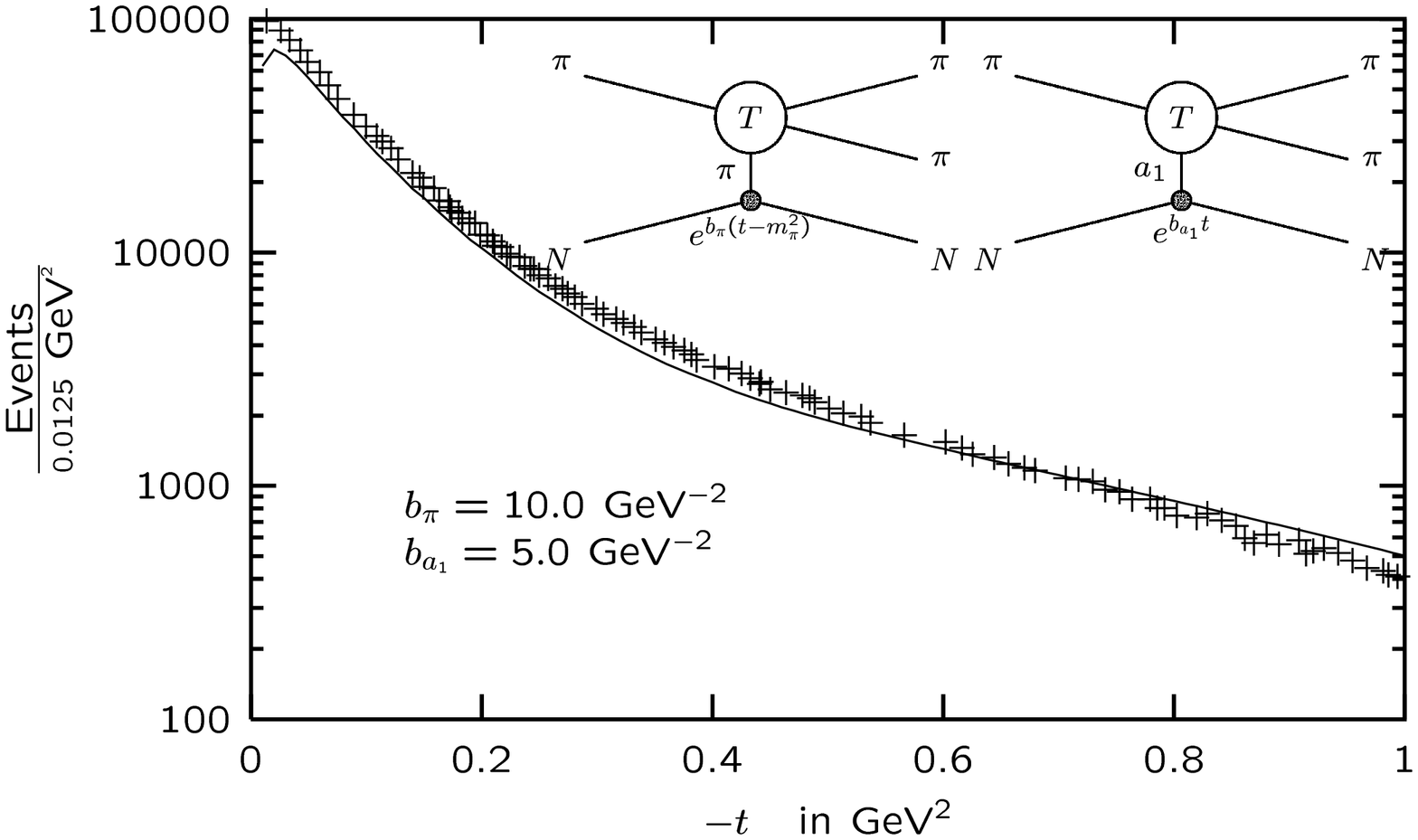}
}}
\caption{\label{dsigmadt}The $\pi^0\pi^0$ production events as a
  function
of the square $t$ of the momentum transfer between proton and neutron.
 These data are used to determine the slope factors b.
Solid line: meson-exchange model including final state interactions
between the produced mesons ( see Feynman diagrams of the insert ).
Crosses: the BNL-E852 data\cite{Gunter:2000am}.}
\end{figure}

For ultrarelativistic beam momenta in the present kinematical regime the reaction
 $\pi^- p \rightarrow \pi^0\pi^0 n$
is a peripheral one. This implies a relatively simple reaction mechanism
which suppresses especially the excitation of nucleon resonances. The 
relevant 
Feynman diagrams are displayed in  Fig.\ref{dsigmadt}. In a peripheral
 reaction one assumes that the incoming pion interacts with the
meson cloud of the proton only once. On the other hand one fully considers the final state interaction between
the produced mesons.
 In  a peripheral charge-exchange reaction, only isovector
mesons have to be considered. The $\rho$-meson cannot contribute because of 
G-parity. This leaves the pion and the $a_1$-meson as
the only relevant mesons with parity $P=(-1)^{J+1}$ to be exchanged in the 
t-channel. The $a_2$ for example cannot contribute in the reaction since it 
has quantum numbers $J^P=2^+$. However the
$a_1$-exchange is known to be important in peripheral $\pi 
N$-reactions\cite{Kaminski:1997gc,Achasov:1998pu}.

The final state interaction of the produced mesons is described by
an improved version of the J\"ulich meson-exchange
model\cite{Lohse:1990ew, Krehl:1997rk}. This means we use the 
Blankenbecler Sugar scattering equation\cite{Blankenbecler:1966gx} 
to generate our pion 
pion $T$-matrix. 
\begin{eqnarray}\nonumber 
&&T_{ij}(\vec k',\vec k;E)=V_{ij}(\vec k',\vec k;E)\\
&& \phantom{T_{ij}} \nonumber +\sum_l \int 
d^3\vec k''V_{il}(\vec k',\vec k'';E) G_l(\vec k'';E)  
T_{lj}(\vec k'',\vec k;E)
\end{eqnarray}
Here $\vec k$ and $\vec k'$ are the momenta of the initial and final 
particles in the center of mass frame and $E$ is the total energy of the 
system. The propagator $G$ has been constructed in a way that ensures 
unitarity for the $S$-matrix and is given by:
\begin{eqnarray}
\nonumber G_l(\vec k;E)&=&\frac{\omega_1(\vec k)+\omega_2(\vec 
k)}{(2\pi)^3 
2\omega_1(\vec k)\omega_2(\vec k)} 
\frac{1}{E^2-(\omega_1(\vec k)+\omega_2(\vec k))^2} 
\end{eqnarray}
with $\omega_{1/2}(\vec k)=\sqrt{\vec k^2+m_{1/2}^2}$. Furthermore $V$ 
is calculated in the one boson exchange approximation including $s$- 
and $t$-channel graphs. The subscripts to 
the transition matrix $T$, the propagator $G$ and the potential $V$ 
indicate 
the coupled channels used in our analysis. They are the $\pi\pi$ and the 
$K\bar K$ channel as well as the newly added $\pi a_1$ reaction channels. 
When adding the latter we used the 
Wess-Zumino Lagrangian\cite{Wess:1967jq} for the $a_1 \rho \pi$-coupling.

We also
investigated whether our $K\bar{K}$-molecule was artifically generated
by the independent choice of $s$- and $t$-channel form factors.
\begin{figure}
\resizebox{0.48\textwidth}{!}{%
 \includegraphics{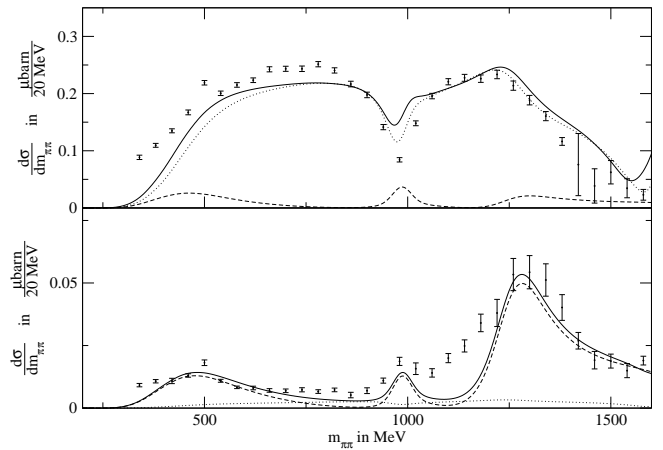}
}
\caption{\label{highlow}
The contribution of the $S$-wave to the total cross section is shown as a function of the
invariant two-pion mass $m_{\pi\pi }$. Solid line: the meson-exchange model;
dotted line: contribution generated by pion exchange at the
proton-neutron vertex;
dashed line: contribution generated by $a_1$ exchange at the
proton-neutron vertex.
In the upper part, the $S$-wave contributions to the cross section
  from \cite{Gunter:2000am} averaged for $0.01<-t<0.1\;\textrm{GeV}^2$
are shown as a function of the invariant two pion mass, while in the
lower part the corresponding data   averaged for $0.4<-t<1.5\;\textrm{GeV}^2$
are shown.
The data are scaled according to the limits given in \cite{Kaminski:2001hv}.}
\end{figure}
Correlating the form factors by
dispersion relations we found no hint in this direction.
In the original model, only one scalar meson $f_0(1400)$ was included.
Now we consider both the $f_0(1370)$ and the $f_0(1500)$ mesons
as $s$-channel diagrams. The couplings of  these mesons to the three reaction
channels considered were adjusted to reproduce the
two-pion decays of the resonances. We found $g_{f_0(1370)K\bar{K}} = 0.551$,
$g_{f_0(1370)\pi a_1} =0.268$,  and
 $g_{f_0(1500)\pi\pi }= g_{f_0(1500)K\bar{K}} =0.188$. These are effective couplings
which also simulate the influence of $4\pi$ decay channels.
This is a minimal extension of the original
J\"ulich model which allows  to discuss the structure of the $f_0(980)$, 
which is our main point of interest.
To analyze  the  decay structure of   the $f_0(1370)$ and the $f_0(1500)$ mesons,
 the inclusion of $4\pi$ decays would be required, however
 \cite{Abele:2001pv}.
 The $\pi\pi$ phase shifts obtained in the new
model are very similar to the  ones of Ref.  \cite{Lohse:1990ew}.

Given the large beam momentum, we describe the initial $\pi$- and $a_1$-meson exchanges
by the corresponding Regge trajectories.
In ultrarelativistic two-pion production reactions, the cross sections decrease
exponentially with the momentum transfer $t$. In the partial wave analysis of the
data, one therefore attaches a slope factor $e^{b_{\pi}(t-m_{\pi}^2)}$.
 The analysis
of the BNL data required the introduction of two different slope factors.
We interprete the two slope factors as effective  form factors of the $pn\pi$-
and the $pna_1$-vertices. Choosing $b_{\pi}=10.0\;\textrm{GeV}^{-2}$ and $b_{a_1}=5.0\;\textrm{GeV}^{-2}$,
the model can reproduce the experimental slope up to
$-t=2\;\textrm{GeV}^2$, see Fig.\ref{dsigmadt}. The full 
$t$-dependence 
is given by:
\begin{eqnarray} \nonumber
\frac{\partial^2\sigma}{\partial m_{\pi\pi}\partial 
t}&=&A_{\pi}\frac{-t}{(t-m_{\pi}^2)^2}e^{b_{\pi}(t-m_{\pi}^2)}\left| 
T_{\pi\pi\rightarrow\pi\pi}(m_{\pi\pi},t)\right|^2\\
\label{prodampli} &&+A_{a_1}(1+tC)^2e^{b_{a_1}t}
\left|T_{\pi a_1\rightarrow\pi\pi}(m_{\pi\pi},t)\right|^2
 \end{eqnarray}
Please note that $A_{\pi}$ and $A_{a_1}$ are not constant and that adding 
the absolute values squared is to account for the helicity structure 
as will be explained later. Furthermore $C$ should be considered a free 
parameter as explained in \cite{Achasov:1998pu} where our value of 
$C=-4.4\textrm{ GeV}^{-2}$ was taken from. 

\begin{figure}
\resizebox{0.48\textwidth}{!}{%
 \includegraphics{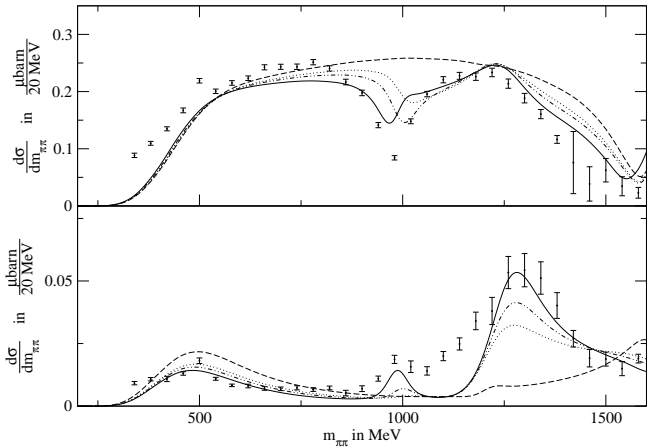}
}
\caption{\label{noKKbar}%
The contribution of the $S$-wave to the total $\pi\pi$ cross section is shown
as a function of the invariant two-pion mass $m_{\pi\pi}$.
The  transition potentials which couple to the
$K\bar{K}$-channel via meson-exchanges in the $t$-channel
are multiplied by a scaling factor,
long-dashed: $\lambda=0.0$,
  dotted:   $\lambda=0.75$,
 dash-dotted:   $\lambda=0.88$,
  solid:   $\lambda=1.0$.
 The data shown are taken from the BNL
  E852 Experiment \cite{Gunter:2000am}.
  The upper and lower part refer to small and large momentum transfers, as in
  Fig.\ref{highlow}.
  }
\end{figure}

In Fig.\ref{highlow}, the $S$-wave contribution to the total cross section
 is shown as a function of the invariant two-pion
mass. In the upper part, the data integrated over the momentum range
$0.01 < -t < 0.1\;\textrm{GeV}^2$ show a broad strength distribution from threshold to
about 1.5 GeV, interrupted by a dip near 980 MeV. 
Our microscopic meson-theoretical model
is able to reproduce this behavior nearly quantitatively. The model
 includes the $\pi\pi$, $K\bar{K}$, and $\pi a_1$ reaction channels, but no coupling to the $\rho\rho$-channel.  For the small momentum transfers displayed
in the upper part of Fig.\ref{highlow},
 the contribution due to the exchange of
a pion  in the initial $t$-channel is dominant.
For invariant masses $m_{\pi\pi}$ ranging from threshold to about 1 GeV,
the experimental $\pi\pi$ phase shifts in the $S$-wave rise
almost linearly  to about $100^{\circ}$. The corresponding partial wave amplitude therefore becomes negative
in the vicinity of  $m_{\pi\pi} = 980\;\textrm{MeV}$. This implies a destructive interference with the amplitude
which describes the $f_0(980)$ meson and generates the dip seen in the data.
At even higher energies the $f_0(1500)$ shows a similar behavior.
 At larger momentum transfers, the broad bump has disappeared in the data and one observes a narrow peak around 1 GeV.
  In that momentum regime (lower part of Fig.\ref{highlow})  the contribution due to the pion in the initial $t$-channel
is negligibly small within our meson exchange model and the $a_1$
 exchange gives the dominating contribution. (This can be traced back
 to the different slope factors.)
 Due to the spin structure interference effects between $a_1$- and
 $\pi$-exchange can be neglected since the $a_1$-emission mainly conserves
 the helicity of the nucleon whereas the $\pi$-emission dominantly
 flips the nucleon helicity. But since the resonant contribution is now
 in phase with the non-resonant background we observe the opposite
 behavior compared to the upper part: the $f_0(980)$ resonance shows
 as a peak.

In contrast to an empirical analysis which 
assumes a  smooth background to which parameterized resonances are added
\cite{Anisovich:1995jy,Achasov:1998pu},
the present approach derives the background in a consistent way within our model.
This is essential for interference effects.
 To illustrate this point, we performed a series of calculations
in which  the transition potentials connecting the $\pi\pi$ channel  and the
 $K\bar{K}$ channel via $t$-channel meson exchanges
 were multiplied by a scaling factor $\lambda$ which
we changed from 0 to 1. The $t$-channel meson exchanges within the $K\bar{K}$
channel were scaled by the same factor. For   $\lambda = 0$, the $\pi\pi$ and  $K\bar{K}$
channel can interact only via $s$-channel diagrams.
The corresponding contributions to the $S$-wave total cross sections are shown
in Fig.\ref{noKKbar}.

For small momentum transfers $t$ between the proton and the neutron (upper part of  Fig.\ref {noKKbar}),
 one finds a broad strength distribution
extending up to 1500 MeV, if the  $t$-channel coupling to the $K\bar{K}$ channel is
switched off ($\lambda=0$).
 Allowing a small
coupling to the $K\bar{K}$ channel  ($\lambda=0.75$),
 the cross section decreases in the energy region
between $1000$ MeV and the onset of the $f_0(1370)$ resonance at about
 $1250$ MeV. As the scaling strength   $\lambda$  is further increased,
a dip develops near $m_{\pi\pi} = 980$ MeV.
For large momentum transfers $t$ between the proton and the neutron (lower part of  Fig.\ref {noKKbar}),
a bump in the vicinity of $m_{\pi\pi} = 980$ MeV appears when the coupling to the
$ K\bar{K}$ channel  is switched on.   

Near  $m_{\pi\pi} = 500$ MeV, a calculation without coupling to
the   $K\bar{K}$ channel overestimates the data. With increasing coupling strength  $\lambda$
the shape of the experimental strength distribution and the relative
size of the bumps centered at $m_{\pi\pi}=500$ MeV and
$m_{\pi\pi}=980$ MeV are reproduced. It is important to 
realize that this feature emerges in
a natural way from a model for the $\pi\pi$ phase shifts when
proper Regge production amplitudes are used. Here for example the 
$(1+tC)^2$ part 
of (\ref{prodampli}), which already appears in early analysis of 
$\pi N\rightarrow\pi\pi N$ scattering data e.g. \cite{Kimel:1977np}, is 
essential to the low energy part of the high momentum transfer case.

\begin{figure}
\resizebox{0.48\textwidth}{!}{%
 \includegraphics{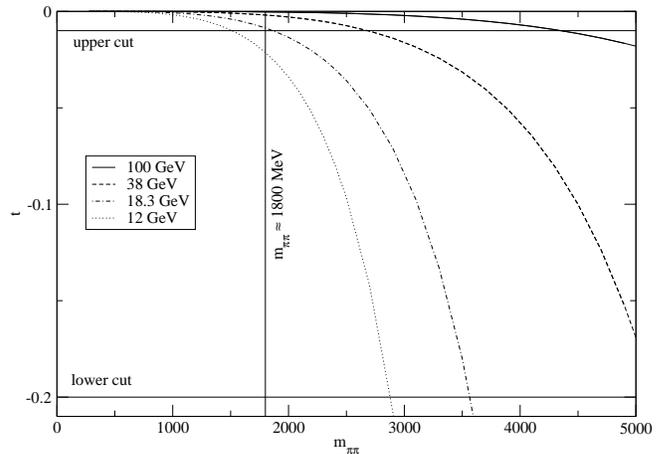}
}
\caption{\label{kinlim}%
The upper kinematic limit of $t$ is plotted against the invariant
mass of the two pion system $m_{\pi\pi}$ for four different beam
energies corresponding to GAMS $100\textrm{ GeV}$, GAMS $38\textrm{
  GeV}$, BNL $18.3\textrm{ GeV}$ and KEK $12\textrm{ GeV}$. 
For comparison also the upper and lower limit applied in the
experimental low momentum transfer cut are shown.
  }
\end{figure}

\begin{figure}
\resizebox{0.48\textwidth}{!}{%
 \includegraphics{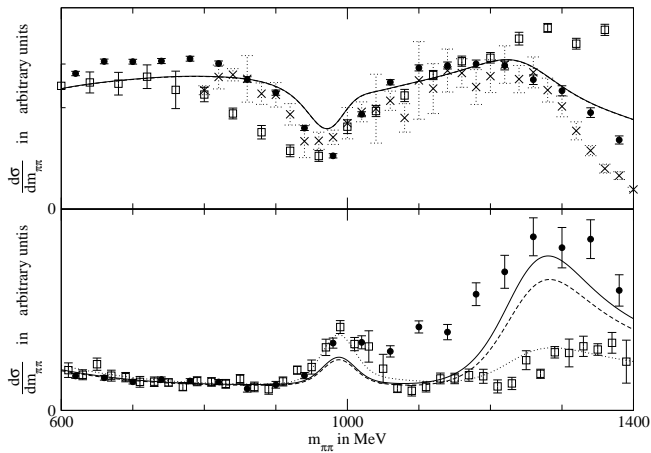}
}
\caption{\label{beamEs}%
The contribution of the $S$-wave to the total $\pi\pi$ cross section
is shown as a function of the invariant two-pion mass
$m_{\pi\pi}$. The upper panel shows our prediction for
$0.01<-t<0.20 \textrm{ GeV}^2$ together with experimental data from
different beam energies. Crosses: $100$ GeV GAMS\cite{Alde:1998mc}, Squares: $38$ GeV GAMS\cite{Alde:1995jj},
Circles: $18.3$ GeV BNL\cite{Gunter:2000am}. The lower panel compares two $t$-ranges and our
corresponding predictions: Circles and solid line: 
$0.3<-t<1.5 \textrm{ GeV}^2$\cite{Gunter:2000am} Squares and interrupted lines: $0.3<-t<1.0
\textrm{ GeV}^2$\cite{Alde:1995jj}. The dashed line is the original
model, the dotted line shows a calculation with $f_0(1370)$ couplings
adjusted to the GAMS data.
  }
\end{figure}

One should further notice that a cross check with data from other experiments with different
$t$-binning and beam energy is of course interesting but can only
contribute limited information in our case. 
To arrive at this conclusion let us have a look at the influence of
the beam energy on the production. Firstly the beam energy enters as a
factor of $\frac{1}{q^2_{beam}s_{tot}}$, which cannot be observed by us, since
the data is not normalized. Secondly it enters by limiting the range
of the $t$-integration. This is shown in Fig. \ref{kinlim} for several
beam energies together with the cut ($0.01<-t<0.2$) applied to the
momentum transfer in the analysis. One observes that for the beam 
energies for which
experimental data is available, the limits on $t$ are essentially determined
by the analysis and not by kinematics if one looks at invariant two
pion masses below $\approx 1800 \textrm{ MeV}$. This is strictly true
for the data under consideration here since we do not consider the
$12\textrm{ GeV}$ KEK data which are only available as an
extrapolation to the pion pole. This means that below an invariant two
pion mass of $1800\textrm{ MeV}$ the data sets should be identical up to an
overall scaling.

The invariant mass range to which our model is applicable extends up
to at most $1.5 \textrm{ GeV}$. This means that the sets of data
stemming from different beam energies should be identical in our invariant
mass region of interest. Thus we compare the two
sets of GAMS data at $38 \textrm{ GeV}$ and at $100 \textrm{ GeV}$
and the $18.3 \textrm{ GeV}$ BNL Data as is shown
in Fig. \ref{beamEs}. The upper panel shows the BNL data (filled
circles) together with the GAMS data at $100 \textrm{ GeV}$
(crosses) and at $38\textrm{ GeV}$ (open squares) for the low momentum
transfer case $0.01<-t<0.20 \textrm{ GeV}^2$. We see that the data
sets and our calculations are up to scaling in good agreement. Only 
the GAMS $38\textrm{ GeV}$ data deviates in shape 
from the other two data sets in the invariant two pion mass region
$0.8 - 1.0 \textrm{ GeV}$ and above $1.2 \textrm{ GeV}$. 

\begin{figure}
\resizebox{0.48\textwidth}{!}{%
 \includegraphics{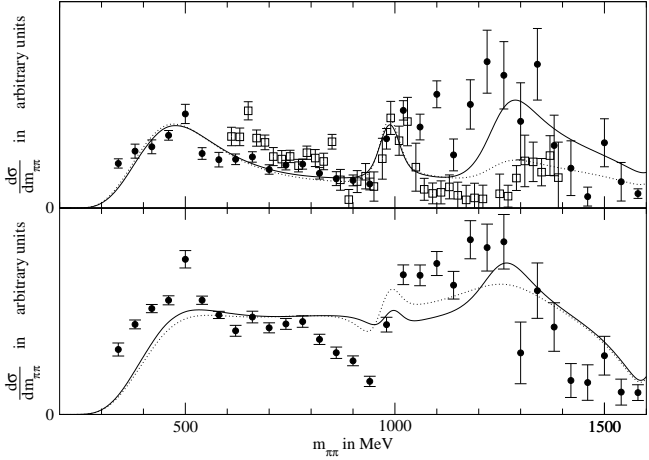}
}
\caption{\label{midt}%
The contribution of the $S$-wave to the total cross section is shown as
a function of the invariant two-pion mass $m_{\pi\pi}$. Both
panels show the BNL data (Circle)\cite{Gunter:2000am} and the GAMS
data (Squares)\cite{Alde:1995jj}. The GAMS data has been derived as
the difference of two data sets and errors have been scaled down by a
factor of four to guide the eye. The curves show the results of our
calculation (solid line) and of a model with the $f_0(1370)$ coupling taken
from the GAMS case in Fig. \ref{beamEs} (dotted line). The upper panel shows momentum
transfers $0.3<-t<0.4\textrm{ GeV}^2$ and the lower panel momentum transfers
$0.2<-t<0.4\textrm{ GeV}^2$ (No data available from GAMS). }
\end{figure}

Comparing the sets of data in the case of high momentum
transfer is not that easy since BNL only quotes a $t$-binning of
$0.4<-t<1.5\textrm{ GeV}^2$ whereas GAMS quotes $0.3<-t<1.0 
\textrm{ GeV}^2$. Nevertheless a comparison might be more rewarding 
since looking at
different $t$-binning means looking at a different ratio of the two
production mechanisms. To have at least a common upper limit we join bins for the BNL case to get a
momentum transfer range of $0.3<-t<1.5 \textrm{ GeV}^2$ and plotted it
together with the GAMS $0.3<-t<1.0 \textrm{ GeV}^2$ data and our
calculations for both $t$-ranges. Having a common upper limit and
knowing that production mainly takes place at low absolute momentum transfers
$|t|$ makes this approach justifiable. This comparison is shown in the 
lower panel of
Fig. \ref{beamEs}. 
The data coincide up to
$1.05\textrm{ GeV}$ from where on they start to deviate strongly.
It is tempting to assign the difference to the production in the
momentum transfer range  $1.0<-t<1.5\textrm{ GeV}^2$ and to interpret
this as the $f_0(1370)$ being a very compact object which can be
produced at large momentum transfers. Our calculation shows a
different behavior: The BNL data is reasonably well described (solid
curve compared to circles), but our model predicts a much too strong
production above $1.05\textrm{ GeV}$ in the case of the GAMS data. The
dashed line is our prediction in the case of the GAMS data which are
shown as squares. In the following section we will point out that the
BNL and GAMS data are inconsistent and that we are not in a position
to judge which one is correct so we also introduce a fit to the GAMS
data (dotted line) by just varying the coupling of the $\pi
a_1$-channel to the $f_0(1370)$ and thus not changing the low momentum
transfer behavior. The good agreement of this second fit to the GAMS
data also at other momentum transfers (as can be seen in Fig. \ref{midt})
demonstrates that our conclusions on the $K\bar{K}$ contribution to the $f_0(980)$
stand firm for both sets of data but only the parameters for the
admixture of the
$f_0(1370)$ need to be questioned. 

To come to the conclusion that the GAMS and BNL Data are inconsistent
we looked at the highest momentum transfer
range where data from both GAMS and BNL are available, 
$0.3<-t<0.4 \textrm{ GeV}^2$. We show these data in the upper panel of 
Fig. \ref{midt}. The GAMS data needed to be derived from the data
published in \cite{Alde:1995jj} by subtracting two sets of data. The
resulting errors have been scaled down by a factor of four so that the
errors shown here roughly correspond to the spreading of the data.
We believe this to be a more realistic estimate of the error. Already at
this momentum transfer range the two sets of data
start to deviate in shape at about $1.05 \textrm{ GeV}$ even though
they should be identical up to a scaling factor. Our calculation again
reproduces the BNL data whereas the GAMS data are
overestimated. When using the coupling parameter for the
$f_0(1370)$ which has been fitted against the high momentum data of GAMS
(dotted lines in Fig. \ref{beamEs} \&  \ref{midt}) instead of the parameter
fitted to the BNL data a good description is achieved, however we find that
our model reproduces the $t$-dependence in the data very well, both if
we look at the BNL data only or at the GAMS data only. Of
course we
cannot resolve the discrepancy between the two data sets. 

A comparison
of our model to the intermediate $t$-range is problematic since our
predictions in this case are very sensitive to the slope parameters for
$\pi$ and $a_1$ exchange, which in turn cannot be fixed to a 
sufficient accuracy by the fit to the $\frac{d\sigma}{dt}$ plot.
This problem arises because our predictions strongly depend
on the point where the production mechanisms become equally important
and this point changes rapidly with the slope parameters. Fitting the data in
this case would mean stronger fine tuning of the parameters than would be
appropriate for a microscopic model like ours. In the lower panel of Fig. \ref{midt} we
nevertheless show our results for this momentum transfer range
($0.2<-t<0.4\textrm{ GeV}^2$). 
In order to demonstrate that there is also a strong variation with the
coupling to the $f_0(1370)$ we show both the calculation with an $f_0(1370)$ as demanded by the BNL
data (solid line) as well as the calculation with the $f_0(1370)$ which has been fitted to the
GAMS data(dotted line). Even though there is no GAMS data available in
this momentum transfer range we can infer from the data shown in the
upper panel in Fig. \ref{midt}, which displays a subset of the
$t$-range shown the lower panel, that above $1.05\textrm{ GeV}$
invariant two pion mass the data points of the GAMS
experiment should be lower than the BNL data. Keeping in mind which size of 
discrepancies has to be
expected between the different experiments we conclude that even for the
medium momentum transfer range our calculation reproduces the main
features of the data as good as one might expect from a microscopic
model like ours.

\begin{figure}
\resizebox{0.48\textwidth}{!}{%
 \includegraphics{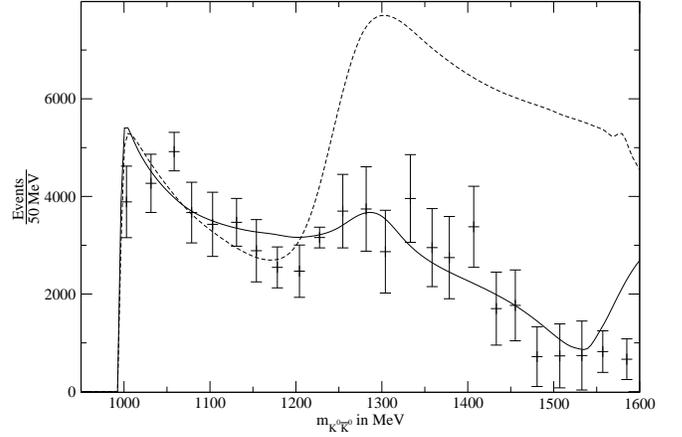}
}
\caption{\label{eventsk0k0}%
The contribution of the $S$-wave to the production $\pi^-p\rightarrow K^0\bar{K}^0n$ is shown
as a function of the invariant two-kaon mass $m_{K\bar{K}}$.
dashed: our model with overestimation due to missing $4\pi$-decays,
solid:  Model with additional $\sigma\sigma$-channel.
The data shown is taken from \cite{Etkin:1982sg} with the bin width of
$50$ MeV not shown.  
  }
\end{figure}

Finally, we compare the results of our model for the reaction $\pi^- p
\rightarrow K^0 \bar{K}^0 n$ with the published data\cite{Etkin:1982sg}.
The model works satisfactorily from threshold up to
about 1200 MeV. Beyond that energy, our model strongly overestimates
the production of neutral kaons (Dashed line in
Fig.\ref{eventsk0k0}). This is understood when comparing our effective
couplings to the decays of the $f_0(1370)$ and $f_0(1500)$ resonances
as listed in \cite{Hagiwara:2002pw}. We used a strong coupling to the $K\bar{K}$
to simulate decays which in reality go to $4\pi$-channels thus naturally
overestimating the kaon production. The solid line in
Fig.\ref{eventsk0k0} shows a version of the model where this shortcoming has been
removed by a mock $\sigma\sigma$-channel to
account for $4\pi$-decays. A good description of the data is obtained
even beyond $1200$ MeV where the partial wave amplitudes are strongly dominated by resonances.

We conclude that the J\"ulich model which predicts a strong $K\bar{K }
$ molecule contribution in the $f_0(980)$ can explain the strong
dependence 
of the
$S$-wave production on the momentum transfer between the proton and
the neutron near $m_{\pi\pi}=980$ MeV by an interference mechanism.

\section{Acknowledgment}
The support of this work by the Deutsche Forschungsgemeinschaft (DFG
447AUS 113/14/0) is gratefully acknowledged.

\bibliography{sassen}

\begin{thebibliography}{25}
\expandafter\ifx\csname natexlab\endcsname\relax\def\natexlab#1{#1}\fi
\expandafter\ifx\csname bibnamefont\endcsname\relax
  \def\bibnamefont#1{#1}\fi
\expandafter\ifx\csname bibfnamefont\endcsname\relax
  \def\bibfnamefont#1{#1}\fi
\expandafter\ifx\csname citenamefont\endcsname\relax
  \def\citenamefont#1{#1}\fi
\expandafter\ifx\csname url\endcsname\relax
  \def\url#1{\texttt{#1}}\fi
\expandafter\ifx\csname urlprefix\endcsname\relax\def\urlprefix{URL }\fi
\providecommand{\bibinfo}[2]{#2}
\providecommand{\eprint}[2][]{\url{#2}}

\bibitem[{\citenamefont{Amsler}(1998)}]{Amsler:1998up}
\bibinfo{author}{\bibfnamefont{C.}~\bibnamefont{Amsler}},
  \bibinfo{journal}{Rev. Mod. Phys.} \textbf{\bibinfo{volume}{70}},
  \bibinfo{pages}{1293} (\bibinfo{year}{1998}),
  \eprint[http://arXiv.org/abs]{hep-ex/9708025}.

\bibitem[{\citenamefont{Amsler}(2002)}]{Amsler:2002ey}
\bibinfo{author}{\bibfnamefont{C.}~\bibnamefont{Amsler}},
  \bibinfo{journal}{Phys. Lett.} \textbf{\bibinfo{volume}{B541}},
  \bibinfo{pages}{22} (\bibinfo{year}{2002}),
  \eprint[http://arXiv.org/abs]{hep-ph/0206 104}.

\bibitem[{\citenamefont{Abele et~al.}(2001)}]{Abele:2001pv}
\bibinfo{author}{\bibfnamefont{A.}~\bibnamefont{Abele}} \bibnamefont{et~al.}
  (\bibinfo{collaboration}{CRYSTAL BARREL}), \bibinfo{journal}{Eur. Phys. J.}
  \textbf{\bibinfo{volume}{C21}}, \bibinfo{pages}{261} (\bibinfo{year}{2001}).

\bibitem[{\citenamefont{Jaffe}(1977)}]{Jaffe:1977ig}
\bibinfo{author}{\bibfnamefont{R.~L.} \bibnamefont{Jaffe}},
  \bibinfo{journal}{Phys. Rev.} \textbf{\bibinfo{volume}{D15}},
  \bibinfo{pages}{267} (\bibinfo{year}{1977}).

\bibitem[{\citenamefont{Weinstein and Isgur}(1990)}]{Weinstein:1990gu}
\bibinfo{author}{\bibfnamefont{J.~D.} \bibnamefont{Weinstein}}
  \bibnamefont{and} \bibinfo{author}{\bibfnamefont{N.}~\bibnamefont{Isgur}},
  \bibinfo{journal}{Phys. Rev.} \textbf{\bibinfo{volume}{D41}},
  \bibinfo{pages}{2236} (\bibinfo{year}{1990}).

\bibitem[{\citenamefont{Lohse et~al.}(1990)\citenamefont{Lohse, Durso, Holinde,
  and Speth}}]{Lohse:1990ew}
\bibinfo{author}{\bibfnamefont{D.}~\bibnamefont{Lohse}},
  \bibinfo{author}{\bibfnamefont{J.~W.} \bibnamefont{Durso}},
  \bibinfo{author}{\bibfnamefont{K.}~\bibnamefont{Holinde}}, \bibnamefont{and}
  \bibinfo{author}{\bibfnamefont{J.}~\bibnamefont{Speth}},
  \bibinfo{journal}{Nucl. Phys.} \textbf{\bibinfo{volume}{A516}},
  \bibinfo{pages}{513} (\bibinfo{year}{1990}).

\bibitem[{\citenamefont{Barnes}(1985)}]{Barnes:1985cy}
\bibinfo{author}{\bibfnamefont{T.}~\bibnamefont{Barnes}},
  \bibinfo{journal}{Phys. Lett.} \textbf{\bibinfo{volume}{B165}},
  \bibinfo{pages}{434} (\bibinfo{year}{1985}).

\bibitem[{\citenamefont{Pichowsky et~al.}(2001)\citenamefont{Pichowsky,
  Szczepaniak, and Londergan}}]{Pichowsky:2001qe}
\bibinfo{author}{\bibfnamefont{M.~A.} \bibnamefont{Pichowsky}},
  \bibinfo{author}{\bibfnamefont{A.}~\bibnamefont{Szczepaniak}},
  \bibnamefont{and} \bibinfo{author}{\bibfnamefont{J.~T.}
  \bibnamefont{Londergan}}, \bibinfo{journal}{Phys. Rev.}
  \textbf{\bibinfo{volume}{D64}}, \bibinfo{pages}{036009}
  (\bibinfo{year}{2001}), \eprint[http://arXiv.org/abs]{nucl-th/0101036}.

\bibitem[{\citenamefont{Oller et~al.}(1998)\citenamefont{Oller, Oset, and
  Pelaez}}]{Oller:1998ng}
\bibinfo{author}{\bibfnamefont{J.~A.} \bibnamefont{Oller}},
  \bibinfo{author}{\bibfnamefont{E.}~\bibnamefont{Oset}}, \bibnamefont{and}
  \bibinfo{author}{\bibfnamefont{J.~R.} \bibnamefont{Pelaez}},
  \bibinfo{journal}{Phys. Rev. Lett.} \textbf{\bibinfo{volume}{80}},
  \bibinfo{pages}{3452} (\bibinfo{year}{1998}),
  \eprint[http://arXiv.org/abs]{hep-ph/9803242}.

\bibitem[{\citenamefont{Achasov and Shestakov}(1998)}]{Achasov:1998pu}
\bibinfo{author}{\bibfnamefont{N.~N.} \bibnamefont{Achasov}} \bibnamefont{and}
  \bibinfo{author}{\bibfnamefont{G.~N.} \bibnamefont{Shestakov}},
  \bibinfo{journal}{Phys. Rev.} \textbf{\bibinfo{volume}{D58}},
  \bibinfo{pages}{054011} (\bibinfo{year}{1998}),
  \eprint[http://arXiv.org/abs]{hep-ph/9802286}.

\bibitem[{\citenamefont{Gunter et~al.}(2001)}]{Gunter:2000am}
\bibinfo{author}{\bibfnamefont{J.}~\bibnamefont{Gunter}} \bibnamefont{et~al.}
  (\bibinfo{collaboration}{E852}), \bibinfo{journal}{Phys. Rev.}
  \textbf{\bibinfo{volume}{D64}}, \bibinfo{pages}{072003}
  (\bibinfo{year}{2001}), \eprint[http://arXiv.org/abs]{hep-ex/0001038}.

\bibitem[{\citenamefont{Alde et~al.}(1995)}]{Alde:1995jj}
\bibinfo{author}{\bibfnamefont{D.}~\bibnamefont{Alde}} \bibnamefont{et~al.}
  (\bibinfo{collaboration}{GAMS}), \bibinfo{journal}{Z. Phys.}
  \textbf{\bibinfo{volume}{C66}}, \bibinfo{pages}{375} (\bibinfo{year}{1995}).

\bibitem[{\citenamefont{Klempt}(2000)}]{Klempt:2000ud}
\bibinfo{author}{\bibfnamefont{E.}~\bibnamefont{Klempt}}
  (\bibinfo{year}{2000}), \eprint[http://arXiv.org/abs]{hep-ex/0101031}.

\bibitem[{\citenamefont{Kondashov}(1999)}]{Kondashov:1998uh}
\bibinfo{author}{\bibfnamefont{A.~A.} \bibnamefont{Kondashov}},
  \bibinfo{journal}{Nucl. Phys. Proc. Suppl.} \textbf{\bibinfo{volume}{74}},
  \bibinfo{pages}{180} (\bibinfo{year}{1999}),
  \eprint[http://arXiv.org/abs]{hep-ph/9811207}.

\bibitem[{\citenamefont{Anisovich et~al.}(1995)\citenamefont{Anisovich,
  Sarantsev, Kondashov, Prokoshkin, and Sadovsky}}]{Anisovich:1995jy}
\bibinfo{author}{\bibfnamefont{V.~V.} \bibnamefont{Anisovich}},
  \bibinfo{author}{\bibfnamefont{A.~V.} \bibnamefont{Sarantsev}},
  \bibinfo{author}{\bibfnamefont{A.~A.} \bibnamefont{Kondashov}},
  \bibinfo{author}{\bibfnamefont{Y.~D.} \bibnamefont{Prokoshkin}},
  \bibnamefont{and} \bibinfo{author}{\bibfnamefont{S.~A.}
  \bibnamefont{Sadovsky}}, \bibinfo{journal}{Phys. Lett.}
  \textbf{\bibinfo{volume}{B355}}, \bibinfo{pages}{363} (\bibinfo{year}{1995}).

\bibitem[{\citenamefont{Anisovich}(2002)}]{Anisovich:2002us}
\bibinfo{author}{\bibfnamefont{V.~V.} \bibnamefont{Anisovich}}
  (\bibinfo{year}{2002}), \eprint[http://arXiv.org/abs]{hep-ph/0208123}.

\bibitem[{\citenamefont{Kaminski et~al.}(1997)\citenamefont{Kaminski, Lesniak,
  and Loiseau}}]{Kaminski:1997gc}
\bibinfo{author}{\bibfnamefont{R.}~\bibnamefont{Kaminski}},
  \bibinfo{author}{\bibfnamefont{L.}~\bibnamefont{Lesniak}}, \bibnamefont{and}
  \bibinfo{author}{\bibfnamefont{B.}~\bibnamefont{Loiseau}},
  \bibinfo{journal}{Phys. Lett.} \textbf{\bibinfo{volume}{B413}},
  \bibinfo{pages}{130} (\bibinfo{year}{1997}),
  \eprint[http://arXiv.org/abs]{hep-ph/9707377}.

\bibitem[{\citenamefont{Krehl et~al.}(1997)\citenamefont{Krehl, Rapp, and
  Speth}}]{Krehl:1997rk}
\bibinfo{author}{\bibfnamefont{O.}~\bibnamefont{Krehl}},
  \bibinfo{author}{\bibfnamefont{R.}~\bibnamefont{Rapp}}, \bibnamefont{and}
  \bibinfo{author}{\bibfnamefont{J.}~\bibnamefont{Speth}},
  \bibinfo{journal}{Phys. Lett.} \textbf{\bibinfo{volume}{B390}},
  \bibinfo{pages}{23} (\bibinfo{year}{1997}),
  \eprint[http://arXiv.org/abs]{nucl-th/9609013}.

\bibitem[{\citenamefont{Blankenbecler and Sugar}(1966)}]{Blankenbecler:1966gx}
\bibinfo{author}{\bibfnamefont{R.}~\bibnamefont{Blankenbecler}}
  \bibnamefont{and} \bibinfo{author}{\bibfnamefont{R.}~\bibnamefont{Sugar}},
  \bibinfo{journal}{Phys. Rev.} \textbf{\bibinfo{volume}{142}},
  \bibinfo{pages}{1051} (\bibinfo{year}{1966}).

\bibitem[{\citenamefont{Wess and Zumino}(1967)}]{Wess:1967jq}
\bibinfo{author}{\bibfnamefont{J.}~\bibnamefont{Wess}} \bibnamefont{and}
  \bibinfo{author}{\bibfnamefont{B.}~\bibnamefont{Zumino}},
  \bibinfo{journal}{Phys. Rev.} \textbf{\bibinfo{volume}{163}},
  \bibinfo{pages}{1727} (\bibinfo{year}{1967}).

\bibitem[{\citenamefont{Kaminski et~al.}(2002)\citenamefont{Kaminski, Lesniak,
  and Rybicki}}]{Kaminski:2001hv}
\bibinfo{author}{\bibfnamefont{R.}~\bibnamefont{Kaminski}},
  \bibinfo{author}{\bibfnamefont{L.}~\bibnamefont{Lesniak}}, \bibnamefont{and}
  \bibinfo{author}{\bibfnamefont{K.}~\bibnamefont{Rybicki}},
  \bibinfo{journal}{Eur. Phys. J. direct} \textbf{\bibinfo{volume}{C4}},
  \bibinfo{pages}{4} (\bibinfo{year}{2002}),
  \eprint[http://arXiv.org/abs]{hep-ph/0109268}.

\bibitem[{\citenamefont{Kimel and Owens}(1977)}]{Kimel:1977np}
\bibinfo{author}{\bibfnamefont{J.}~\bibnamefont{Kimel}} \bibnamefont{and}
  \bibinfo{author}{\bibfnamefont{J.}~\bibnamefont{Owens}},
  \bibinfo{journal}{Nucl. Phys.} \textbf{\bibinfo{volume}{B122}},
  \bibinfo{pages}{464} (\bibinfo{year}{1977}).

\bibitem[{\citenamefont{Alde et~al.}(1998)}]{Alde:1998mc}
\bibinfo{author}{\bibfnamefont{D.}~\bibnamefont{Alde}} \bibnamefont{et~al.}
  (\bibinfo{collaboration}{GAMS}), \bibinfo{journal}{Eur. Phys. J.}
  \textbf{\bibinfo{volume}{A3}}, \bibinfo{pages}{361} (\bibinfo{year}{1998}).

\bibitem[{\citenamefont{Etkin et~al.}(1982)}]{Etkin:1982sg}
\bibinfo{author}{\bibfnamefont{A.}~\bibnamefont{Etkin}} \bibnamefont{et~al.},
  \bibinfo{journal}{Phys. Rev.} \textbf{\bibinfo{volume}{D25}},
  \bibinfo{pages}{1786} (\bibinfo{year}{1982}).

\bibitem[{\citenamefont{Hagiwara et~al.}(2002)}]{Hagiwara:2002pw}
\bibinfo{author}{\bibfnamefont{K.}~\bibnamefont{Hagiwara}} \bibnamefont{et~al.}
  (\bibinfo{collaboration}{Particle Data Group}), \bibinfo{journal}{Phys. Rev.}
  \textbf{\bibinfo{volume}{D66}}, \bibinfo{pages}{010001}
  (\bibinfo{year}{2002}).

\end{thebibliography}
\end{document}